\documentclass[twocolumn]{article}

\usepackage{PRIMEarxiv}

\usepackage[utf8]{inputenc} 
\usepackage[T1]{fontenc}    
\usepackage{hyperref}       
\usepackage{url}            
\usepackage{booktabs}       
\usepackage{amsfonts}       
\usepackage{nicefrac}       
\usepackage{microtype}      
\usepackage{lipsum}
\usepackage{fancyhdr}       
\usepackage{graphicx}       
\usepackage{amsmath}
\graphicspath{{media/}}     

\title{A High-Frequency Flexible Ultrasonic Cuff Implant for High-Precision Vagus Nerve Ultrasound Neuromodulation
\thanks{This work was supported in part by the ECSEL Joint Undertaking project Moore4Medical, grant number H2020-ECSEL-2019IA-876190.}
}

\author{
  Cornelis van Damme \\
 \textit{Department of Microelectronics} \\
\textit{Delft University of Technology}\\
Delft, The Netherlands \\
  \texttt{nvdamme@hotmail.nl} \\
   \And
  Gandhika K. Wardhana \\
  \textit{Department of Microelectronics} \\
\textit{Delft University of Technology}\\
Delft, The Netherlands \\
  \texttt{g.k.wardhana@tudelft.nl} \\
    \And
  Andrada Iulia Velea \\
   \textit{Department of Microelectronics} \\
\textit{Delft University of Technology}\\
Delft, The Netherlands \\
\\
   \textit{Department of System Integration} \\ 
   \textit{and Interconnection Technologies} \\
\textit{Fraunhofer IZM}\\
 Berlin, Germany \\
  \texttt{a.velea-1@tudelft.nl} \\
    \And
  Vasiliki Giagka$^{**}$\\
 \textit{Department of Microelectronics} \\
\textit{Delft University of Technology}\\
Delft, The Netherlands \\
\\
\textit{Department of System Integration} \\
\textit{and Interconnection Technologies} \\
\textit{Fraunhofer IZM}\\
 Berlin, Germany \\
  \texttt{v.giagka@tudelft.nl} \\
    \And
  Tiago L. Costa$^{**}$ \\
  \textit{Department of Microelectronics} \\
\textit{Delft University of Technology}\\
Delft, The Netherlands \\
  \texttt{t.m.l.dacosta@tudelft.nl} \\
\\
  $^{**}$These authors contributed equally
}

\begin{document}
\twocolumn[
\begin{@twocolumnfalse}
\maketitle

\begin{abstract}
In the emerging research field of bioelectronic medicine, it has been indicated that neuromodulation of the Vagus Nerve (VN) has the potential to treat various conditions such as epilepsy, depression, and autoimmune diseases. In order to reduce side effects, as well as to increase the effectiveness of the delivered therapy, sub-fascicle stimulation specificity is required. In the electrical domain, increasing spatial selectivity can only be achieved using invasive and potentially damaging approaches like compressive forces or nerve penetration. To avoid these invasive methods, while obtaining a high spatial selectivity, a 2~mm diameter extraneural cuff-shaped proof-of-concept design with integrated Lead Zirconate Titanate (PZT) based ultrasound (US) transducers is proposed in this paper. For the development of the proposed concept, wafer-level microfabrication techniques are employed. 
Moreover, acoustic measurements are performed on the device, in order to characterize the ultrasonic beam profiles of the integrated PZT-based US transducers. A focal spot size of around 200~$\mu m$ by 200~$\mu m$ is measured for the proposed cuff. Moreover, the curvature of the device leads to constructive interference of the US waves originating from multiple PZT-based US transducers, which in turn leads to an increase of 45\% in focal pressure compared to the focal pressure of a single PZT-based US transducer. Integrating PZT-based US transducers in an extraneural cuff-shaped design has the potential to achieve high-precision US neuromodulation of the Vagus Nerve without requiring intraneural implantation.
\end{abstract}

\vspace{10pt}
\textit{\footnotesize $^*$This work was supported in part by the ECSEL Joint Undertaking project Moore4Medical, grant number H2020-ECSEL-2019IA-876190.}
\end{@twocolumnfalse}
]

\section{Introduction}
The application of ultrasound (US) technologies in the medical field has been extended from diagnostic imaging to therapeutic neuromodulation \cite{Oluigbo2011, Blackmore2019, Kamimura2020}. Among several stimulation targets, Vagus Nerve Stimulation (VNS) by means of focused US has been explored in recent years \cite{Downs2018, Kim2020, Cotero2020, Kawasaki2019}. The Vagus Nerve (VN) is a cranial nerve, part of the parasympathetic nervous system, consisting of afferent and efferent neurons \cite{Yuan2016a, Megan2019, Naveen2023, Duncan2005}. The VN nerve fascicles comprise of different nerve fibers, classified, according to Erlanger Gasser as type A, B, and C, each having their own functions, sizes (ranging from $<$0.5~$\mu m$ up to 10~$\mu m$), and conduction velocities (ranging from 0.5 to 120~m/s) \cite{Yuan2016a, Megan2019}. The VN is involved in the autonomic, cardiovascular, respiratory, gastrointestinal, immune, and endocrine systems \cite{Yuan2016a}. Research shows that stimulation is useful in the therapy of epilepsy, depression, and several chronic diseases like Alzheimer’s disease, anxiety, congestive heart failure, pain, tinnitus, and inflammatory diseases \cite{Yuan2016a, Duncan2005, Johnson2018, Panebianco2022, Howland2014,  Lee2017, Yuan2016b}. Targeted stimulation on the sub-fascicle level is needed since unintended stimulation of other fascicles can lead to severe side effects \cite{Yuan2016b, Dirr2019}. 
\begin{figure}[t!]
    \centering
    \includegraphics[width=0.5\textwidth]{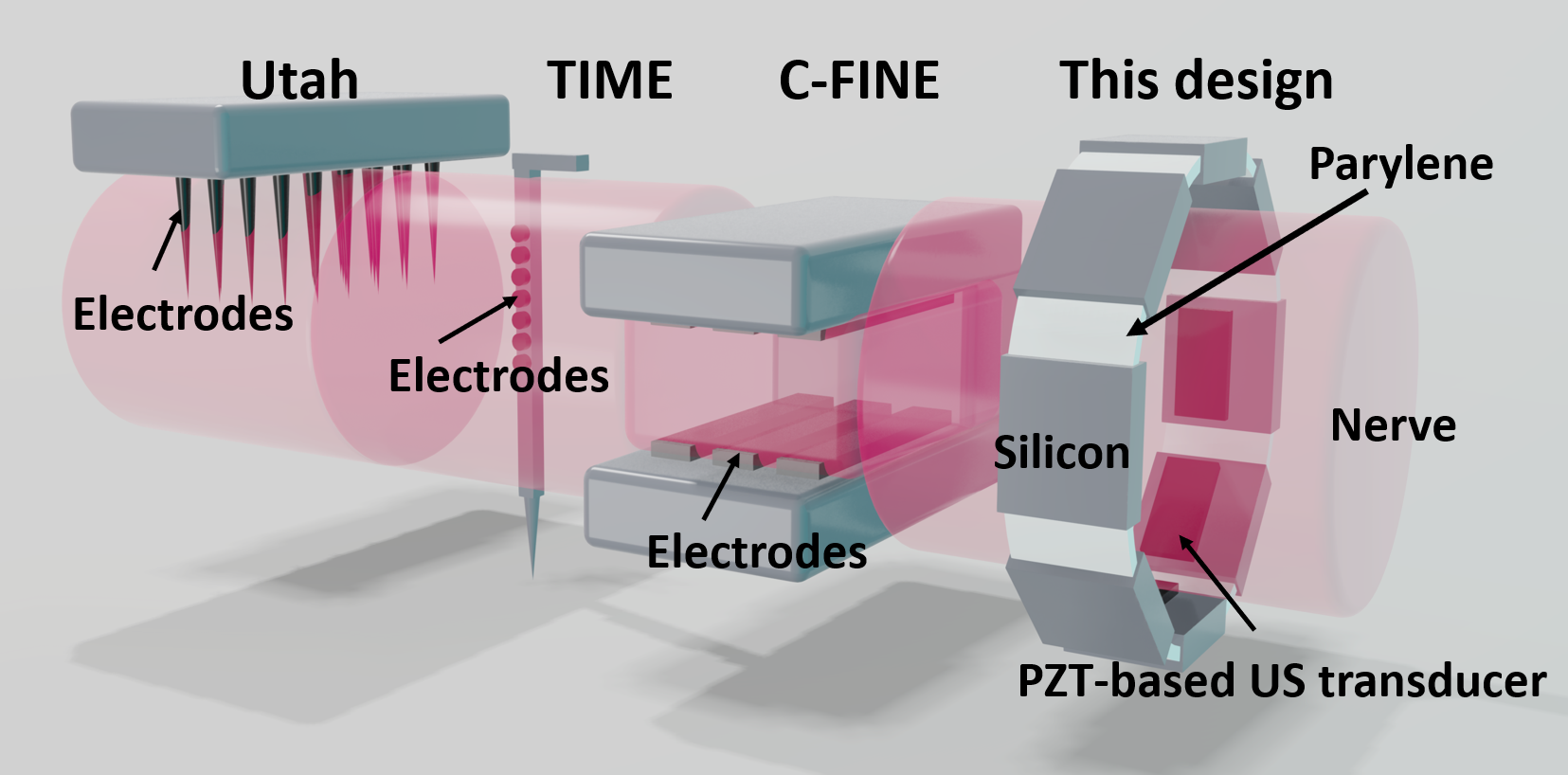}
    \caption{A comparison of the Utah, TIME, and C-FINE electrodes with the current design.}
    \label{fig:fig_concept}
\end{figure}  
Conventionally, electricity is used to interact with the peripheral nervous system \cite{Giagka2018}. 
Transcutaneous VNS (tVNS) has been proposed as a non-invasive method \cite{Yuan2016b, Yap2020, Smet2021, Shin2023, Levitsky2020, Gurel2018}. Although studies show that activation is elicited, the envisioned sub-fascicle stimulation resolution is not met \cite{Pashaei2020}. Improved resolution can be achieved with implantable devices having embedded electrodes. A promising type of electrode for stimulation is the cuff electrode \cite{Rijnbeek2018,  Rodriguez2000, Stieglitz2000, Forssell2019, Haugland1997}. To reach the sub-fascicle resolution with electrodes, techniques like composite flat interface nerve electrodes (C-FINE) \cite{Freeberg2017}, slowly penetrating inter-fascicular nerve electrodes (SPINE) \cite{Tyler1997} and intra-fascicular techniques like longitudinal intra-fascicular electrode (LIFE) \cite{Rijnbeek2018} and transverse intra-fascicular multichannel electrode (TIME) \cite{Rijnbeek2018} and microelectrode arrays (MEAs), for example, the Utah array \cite{Kim2018}, are being developed \cite{Yildiz2020}. The disadvantages of the aforementioned techniques are the needed compressive force and high invasiveness which increase the risk of damage to the nerve during implantation (Fig.~\ref{fig:fig_concept}). Therefore, this makes these techniques unsuitable for chronic applications. \\
\begin{figure}[b!]
    \centering
    \includegraphics[width=0.5\textwidth]{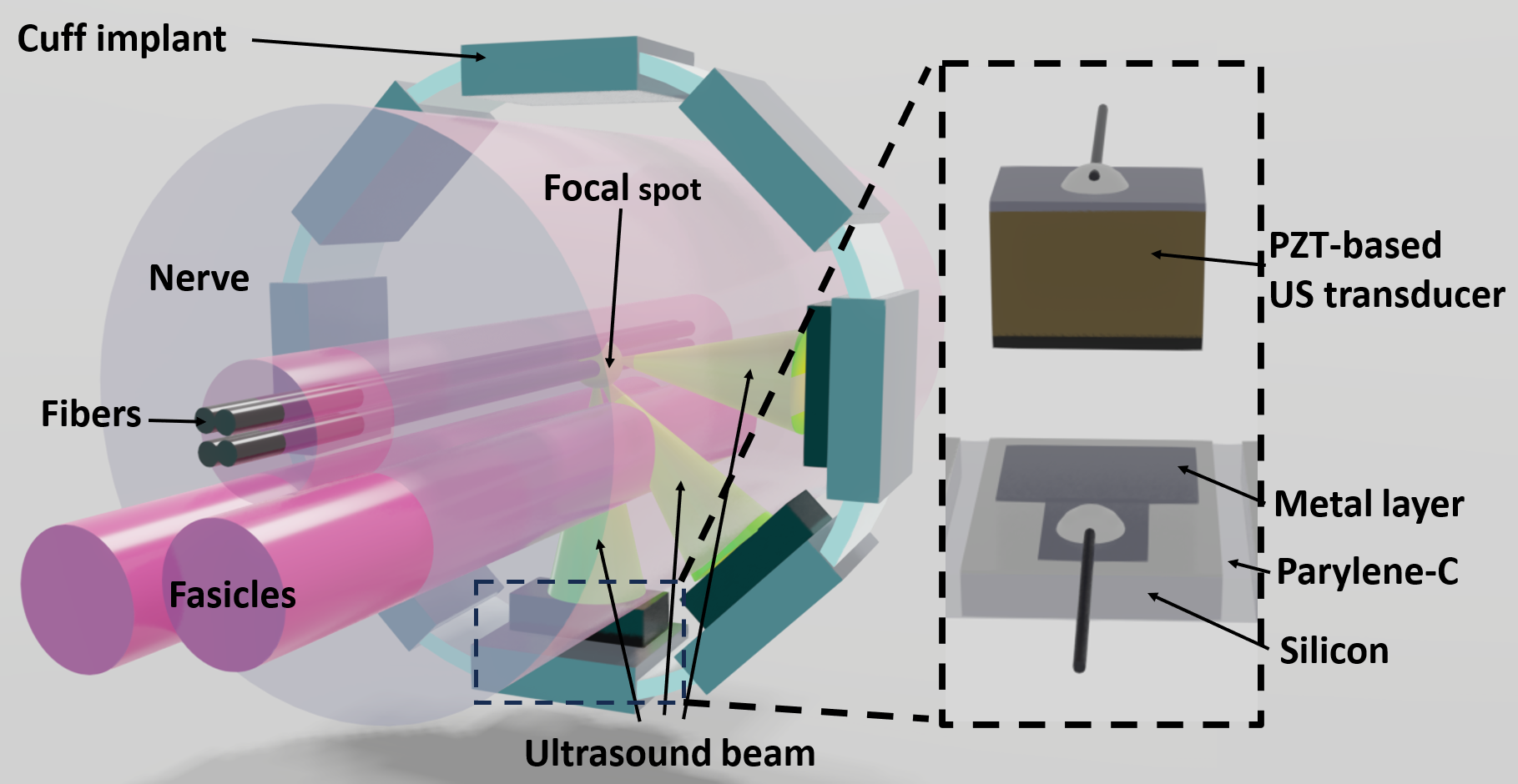}
    \caption{Illustration of the proposed cuff implant concept for high-precision VN
acoustic stimulation. A zoomed-in representation of the comprising elements
and the airbacking layer is also shown.}
    \label{fig:fig_first}
\end{figure} 
Instead of using electrodes, integrating Lead Zirconate Titanate (PZT) based US transducers in a cuff implant form factor would enable the possibility of delivering US neuromodulation, extraneurally, yet with a high spatial resolution (Fig.~\ref{fig:fig_concept}). It has been previously demonstrated that a focal spot of 110~$\mu m$ by 570~$\mu m$ can be achieved when capacitive micromachined ultrasound transducers (CMUTs) are placed under the nerve and are geometrically curved at radii matching that of the VN \cite{Kawasaki2019}. Based on the well-described physical phenomena of US, it has been shown that US can be beam steered \cite{Costa2021, kawasaki2021} and can propagate through tissue for several centimeters without causing damage and side effects \cite{Blackmore2019, Kamimura2020, Obrien2007, Kele2012}. Despite the biological mechanisms of US neuromodulation not yet being perfectly understood, it is likely that different combinations of partially overlapping mechanisms occur in the cell membrane depending on the US pulse regime \cite{Blackmore2019, Kamimura2020, Downs2018, Plaksin2014, Heimburg2005, Oh2019, Colucci2009}. Several studies show that focused US can elicit a physiological response in nerves \cite{Downs2018, Kim2020, Cotero2020, Forssell2019, Colucci2009, Lee2009, Juan2014,  Foley2008}. \\
US waves are generated by either bulk piezoelectric transducers, or by flexural mode transducers, such as CMUTs and piezoelectric micromachined ultrasound transducers (PMUTs) \cite{Costa2021, Rathod2020, Shen2020, Wang2015, Yang2013}. For bulk mode PZT-based US transducers, which are characterized by a high transmit electroacoustic sensitivity ($S_{tx}=\frac{P_{peak}}{V_{driving}}$, where $P_{peak}$ is the peak output pressure [kPa] and $V_{driving}$ the driving voltage [V]) and a high-quality factor \cite{Costa2021}, PZT ceramics are commonly used due to their superior piezoelectric constants \cite{Akasheh2003}. These are important characteristics for US neuromodulation as they lead to higher and more stable pressure amplitudes per driving voltage \cite{Costa2021, Rathod2020, Shen2020}. PMUT and CMUT devices have a lower $S_{tx}$ and lower quality factor, and hence are more suitable for high-quality imaging and sensing applications where bandwidth is important \cite{Costa2021, Akasheh2003}. The pressure output of an integrated CMUT-array in a cuff implant form factor using 25~$V_{pp}$ for excitation with beam steering has been measured to generate 1.7~MPa at most ($S_{tx}$ = 68~kPa/V) \cite{kawasaki2021}. Another planar design with a 2D PZT-based US transducer array with 5~$V_{pp}$ generated up to 0.1~MPa ($S_{tx}$ = 20~kPa/V) \cite{Costa2021}. As the output pressure correlates with the driving voltage \cite{Rivandi2023} and the focusing of the beam, the $S_{tx}$ is a good parameter for comparison. \\
Currently, there is no consensus on the amount of intensity or pressure needed for neuromodulation of peripheral nerves. However, research suggests that peripheral nerves require higher pressures than e.g. brain tissue for neuromodulation and that pressures in the range of 3~MPa are sufficient\cite{Downs2018}. To date, a method to integrate bulk PZT-based US transducers in a flexible cuff compatible with VNS was not yet demonstrated \cite{Pashaei2020}. \\
In this paper, a form factor compatible with the VN with integrated PZT-based US transducers is proposed (Fig.~\ref{fig:fig_first}). We investigate whether this design can reach high acoustic pressures with low peak-to-peak driving voltages and still maintain high spatial resolution. The organization of the paper is as follows: in section \ref{sec:sim} the design choices and the necessary COMSOL Multiphysics \cite{Comsol} simulations are elaborated upon. Section~\ref{sec:design} describes the design and elaborates on the wafer-level microfabrication process flow and the assembly of the PZT-based cuff prototypes (section~\ref{subsec:fab}). In section~\ref{sec:char} the device is characterized and the acoustic measurements are described. The results are discussed in section~\ref{sec:disc}, whereas section \ref{sec:concl} draws the conclusions. 

\section{Simulations}
\label{sec:sim}
\begin{table}[b!]
\centering
\normalsize
\caption{Design parameters}
\begin{tabular}{l|l}
\hline\hline
\textbf{Parameters} & \textbf{Numbers} \\
\hline
\textbf{Frequency $[MHz]$}          & 8.4    \\  
\textbf{Diameter $[mm]$}            & 2  \\ 
\textbf{\# of PZT-based US transducers}            & 3  \\ 
\textbf{PZT thickness $[\mu m]$} & 254  \\  
\textbf{PZT aperture $[\mu m]$}      & 1100  \\ 
\textbf{PZT length $[\mu m]$}      & 450  \\ 
\textbf{Focal length $[mm]$}         & 1  \\     
\textbf{Active area $[mm^2]$}        & 1.49 \\ 
\hline \hline
\end{tabular}
\label{tab:design_param}
\end{table}

\begin{figure*}[t!]
   \centering
    \includegraphics[width=\textwidth]{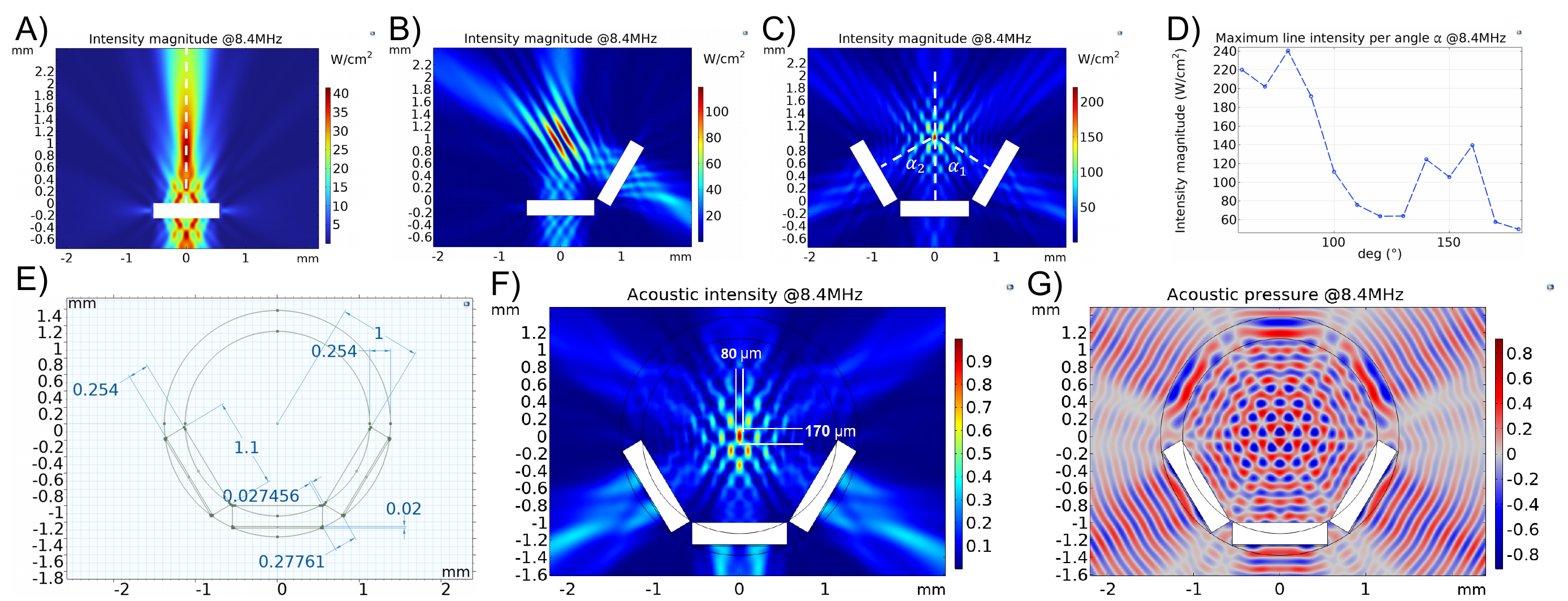}
\caption{Simulation results A), B), and C) show the focal spot size and acoustic intensity for one, two, and three PZT-based US transducers in a curved configuration. D) shows the relation between the angle between three PZT-based US transducers and the maximum acoustic intensity magnitude. E) shows the dimensions for the simulated cuff design, whereas F) and G) show the acoustic intensity magnitude and pressure profiles respectively. }
   \label{fig:sim1}
\end{figure*} 

The concept, shown in Fig.~\ref{fig:fig_concept}, is a cuff-shaped,
island-bridge structure with three 8.4~MHz PZT-based US transducers. In Table~\ref{tab:design_param} the main design parameters are given. The inner diameter of the cuff is 2~mm, as the VN has a diameter of about 2-4~mm \cite{Kawasaki2019}. The aperture of the PZT relates to the focal length and driving frequency according to \cite{Gougheri2019}:
\begin{equation}
    N = \frac{f_pL^2}{4v}
    \label{eq:focal_length}
\end{equation}
where $N$ is the focal length [m], $L$ the aperture [m], $f_p$ the driving frequency [Hz] and $v$ the speed of sound in the medium [m/s]. The focal length of each PZT-based US transducer has been designed to be around 1~mm, such that the focal point of all PZT-based US transducers comprising the cuff is in the center of the design, as well as, of the nerve. 
As the cuff form factor is defined with a radius of 1~mm, the aperture of single PZTs can also not be larger than the chord of 12.5~\% of the circumference, otherwise, it will limit the circular shape. \\
Frequencies for neuromodulation in pre-clinical or clinical research can scale from sub-MHz (transcranial US neuromodulation) to a few MHz (VNS). Increasing the frequency leads to a tradeoff between spatial resolution and absorption, hence the frequency should be carefully set. The driving frequency is inversely proportional to the aperture \eqref{eq:focal_length}, the focal spot size \eqref{eq:FWHM} and \eqref{eq:DOF}, and the thickness \eqref{eq:PZT_thickness} of the PZT \cite{Rathod2020, Gougheri2019}. The equations for the full width at half bandwidth \textit{FHWM} \cite{Costa2021}, the depth of field \textit{DOF} \cite{Costa2021}, and thickness of the PZT at resonance ($t_{PZT}$) \cite{Akasheh2003} are given in \eqref{eq:FWHM}, \eqref{eq:DOF}, and \eqref{eq:PZT_thickness} respectively. 
\begin{equation}
    FWHM \propto \frac{\lambda Z_m}{L}
    \label{eq:FWHM}
\end{equation}
where $\lambda$ is the wavelength of the US waves [m], $Z_m$ is the focal depth [m] and $L$ the aperture [m]. 

\begin{equation}
    DOF \propto \frac{\lambda Z_m^2}{L^2}
    \label{eq:DOF}
\end{equation}

\begin{equation}
    t_{PZT} = \frac{\lambda}{2}
    \label{eq:PZT_thickness}
\end{equation}

 In this study, it has been assumed that the acoustic wave is propagating in a homogeneous medium and that there is no gap between the implant and the nerve. The PZT thickness, defining the resonance frequency, can constrain the curvature of the design as the top of the PZTs could touch each other for large PZT thicknesses. As the thickness of the PZTs is in the range of the silicon thickness (around 300~$\mu m$), it does not constrain the design. Moreover, the frequency determines the aperture, whereas the aperture has a tradeoff between the focal length and the maximum size for curvature. Therefore, the frequency should be set to be as high as possible to have a high spatial resolution, yet with the PZT-based US transducer size fitting the design dimensions. Therefore, a frequency of 8.4~MHz has been set. Moreover, other research shows that similar driving frequencies (9.56 and 8.4~MHz) provide resolution in the \mbox{$\mu m$-range} \cite{Costa2021, Gougheri2019}.

\subsection{Methods}
\label{subsec:sim_methods}
To define the effect of the number of PZT-based US transducers and to verify the design, COMSOL Multiphysics simulations have been performed. The 2D finite element method simulations have been conducted in the frequency domain, using the pressure acoustics, solid mechanics, and electrostatics COMSOL models. A free triangular mesh with a maximum element size of $v/f_{p}/8$ has been used. Water medium has been used as a replacement for nerve tissue since the acoustic properties are similar \cite{Rathod2020}. The boundary of the water medium is set to be perfectly matched to avoid reflections at the edges. In addition, PZT-5H has been used as a piezoelectric material for the PZT-based US transducers and a driving voltage of 10~$V_{pp}$ has been defined, being the maximum output voltage of the function generator used during measurements (Section~\ref{sec:char}). To ensure the focal point is in the center of the device, the distance between the surface of a PZT-based US transducer and the center has been set to 1~mm.  \\
The first simulation has been done to investigate the effect of the number of PZT-based US transducers on the acoustic profile and pressure levels. The number of PZT-based US transducers has been swept from an individual PZT-based US transducer to three PZT-based US transducers. The next simulation is a rotational sweep of the angles $\alpha _1$, $\alpha _2$  which are the angles between one of the side PZT-based US transducers and the bottom-middle PZT-based US transducer (Fig.~\ref{fig:sim1}C). These angles are equal for the left and right side ($\alpha _1 = \alpha _2$) and are being swept from 50$^\circ$ to 180$^\circ$. \\
Later, a full design with three PZT-based US transducers and a polymer ring of parylene-C was simulated to verify the focal spot and the design dimensions. The cuff implant form factor has been modeled as a perfect circle. The simulation dimensions are shown in Fig.~\ref{fig:sim1}E.

\subsection{Results}
\label{subsec:sim_results}
 From simulations, it was found that the increase in the number of PZT-based US transducers increases the acoustic intensity magnitude in case they are placed in a curved configuration. According to the simulations, the focal intensity is 40~$W/cm^2$ for one PZT-based US transducer and increases to 120~$W/cm^2$ for two PZT-based US transducers and to 210~$W/cm^2$ for three PZT-based US transducers (Fig.~\ref{fig:sim1}A, Fig.~\ref{fig:sim1}B, and Fig.~\ref{fig:sim1}C) respectively. The increase of the acoustic intensity becomes less significant the more PZT-based US transducers are added. Moreover, with more PZT-based US transducers the focal spot size decreases and destructive interference patterns appear. \\
The result of a sweep of three PZT-based US transducers is shown in Fig.~\ref{fig:sim1}D. It can be observed that for smaller angles between the PZT-based US transducers, a higher acoustic intensity magnitude exists in this form factor. The intensity for an angle of 180$^\circ$, so the PZT-based US transducers oppose each other, is reduced to 75\% of the maximum acoustic intensity magnitude. Although opposite-placed PZT-based US transducers might be beneficial in different designs and in cases of beam steering, in this design, it has been concluded based on this simulation that opposite-placed PZT-based US transducers should be avoided. This limits the placement of PZT-based US transducers to only 40\% of the cuff circumference. \\
Moreover, the number of PZT-based US transducers is determined by the aperture of the PZTs and the inter-PZT distance. The aperture of the PZTs is set by the aforementioned driving frequency. The inter-PZT distance between the PZT-based US transducers when curved, is optimized to be a multiple of $\lambda _{water}/2$ for minimizing the side lobes while having the smallest distance (Fig.~\ref{fig:sim1}D). For a driving frequency of 8.4~MHz, three PZT-based US transducers do fit in the 2~mm cuff design (Fig.~\ref{fig:sim1}E). 
\begin{figure}[!b]
    \centering
    \includegraphics[width=0.49\textwidth]{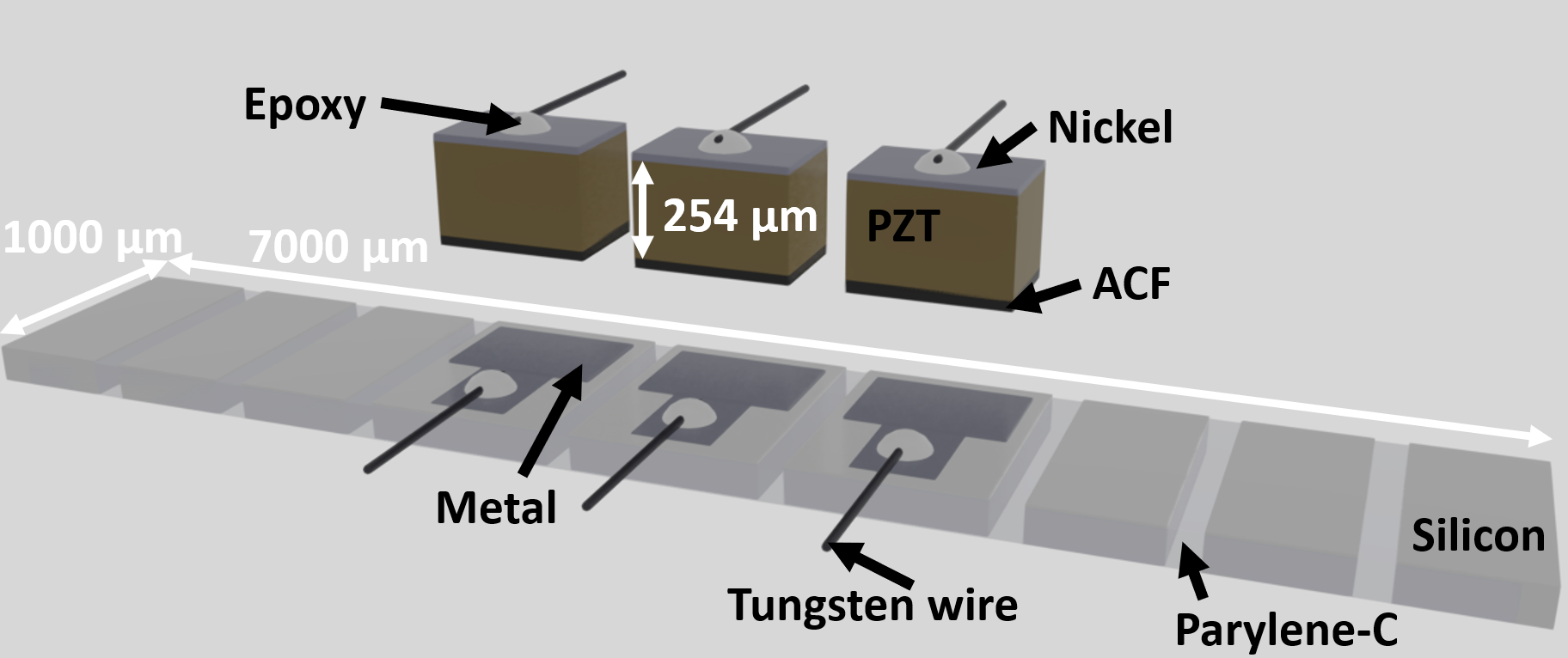}
    \caption{Planar design of the proposed cuff implant.}
    \label{fig:design}
\end{figure} 
\begin{figure*}[t!]
    \centering
    \includegraphics[width=\textwidth]{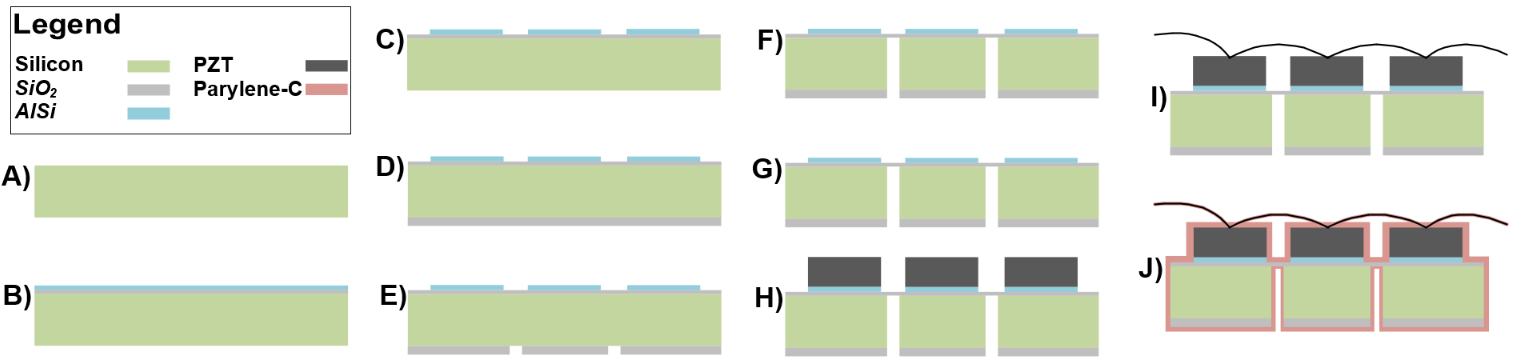}
    \caption{The wafer-level microfabrication process steps A) the process starts with a wafer on which alignment markers are etched B) a layer of $SiO_2$ and on top a layer of $AlSi$ is deposited C) the $AlSi$ layer is patterned D) a $SiO_2$ layer is deposited at the backside E) the $SiO_2$ at the backside is patterned F) the silicon substrate is etched from the backside G) the wafer is diced H) PZTs are placed on top I) tungsten wire is attached on top of the PZTs J) the device is encapsulated with parylene-C.}
    \label{fig:processing}
\end{figure*} 
The acoustic intensity magnitude and pressure profiles for the cuff implant design can be found in Fig.~\ref{fig:sim1}F and \ref{fig:sim1}G, respectively. The acoustic waves are emitted from both the front- and back-side of the PZT-based US transducer. 
Note that in COMSOL the intensity is a vector whereas the pressure is a scalar, resulting in different profiles. It can be observed that the focal spot for the acoustic intensity has a size of 80~$\mu m$ by 170~$\mu m$ and it is located in the center of the cuff shape.

\section{Design}
\label{sec:design}
The development of the proposed cuff is based on wafer-level microfabrication processes \cite{Costa2021, Shi2018, Shi2021, Shi2020}. The flexibility of the final device is provided by the island-bridge approach where silicon islands are etched and interconnected with each other via a parylene-C layer. 
The metal layer provides the electrical connection to the PZT-based US transducers (Fig.~\ref{fig:design}). The contact pads (500~$\mu m$ by 500~$\mu m$) are directly connected to this metal layer with 500~$\mu m$-width traces. \\ 
The single planar device is 7~mm by 1~mm (Fig.~\ref{fig:design}). According to the simulations in section~\ref{sec:sim}, the resonance frequency and thus, the driving frequency of the cuff concept is 8.4~MHz resulting in a PZT thickness of around 254~$\mu m$. Taking the design constraints into account, only three PZT-based US transducers can be placed (Fig.~\ref{fig:design}). The sizes of the PZT-based US transducers can be found in table~\ref{tab:design_param}. 
\begin{figure}[!b]
    \centering
    \includegraphics[width=0.5\textwidth]{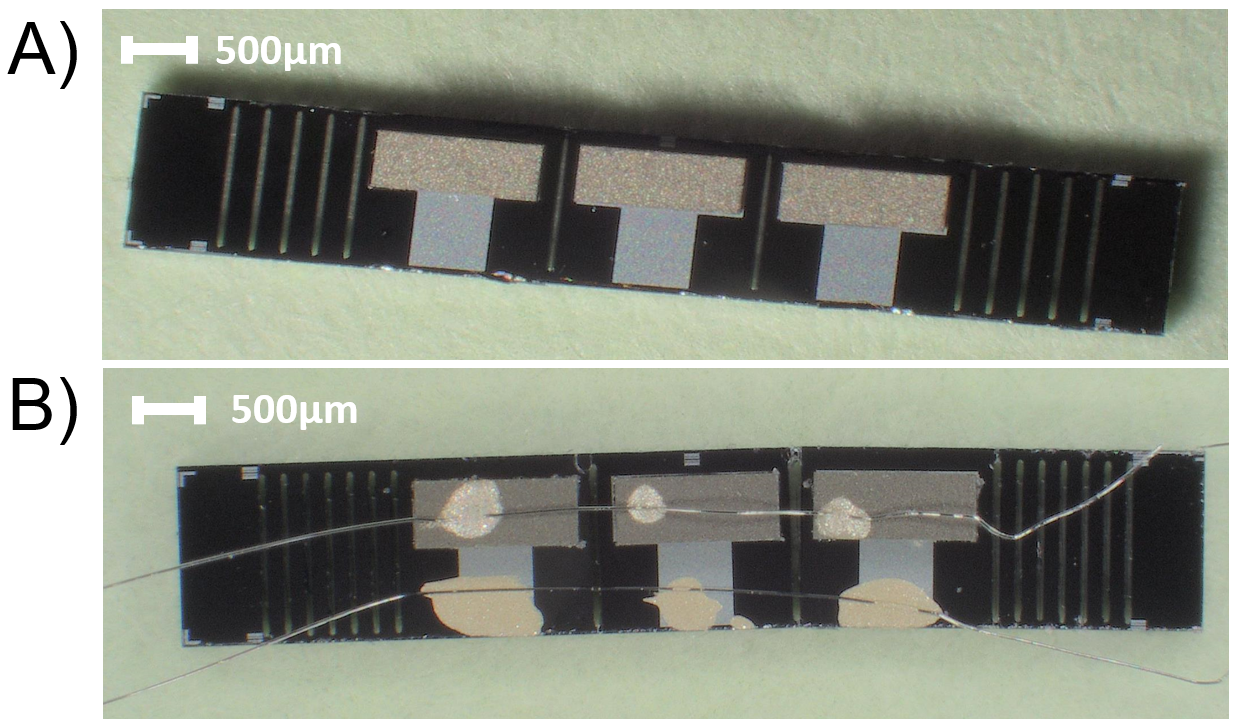}
    \caption{The design and fabricated device. A) is a microscopic topview of a single, microfabricated device after step H, whereas B) is a micrograph of the fabricated device after the attachment of tungsten wire (step I).}
    \label{fig:fig_devices}
\end{figure} 
\begin{figure*}[h!]
    \centering
    \includegraphics[width=\textwidth]{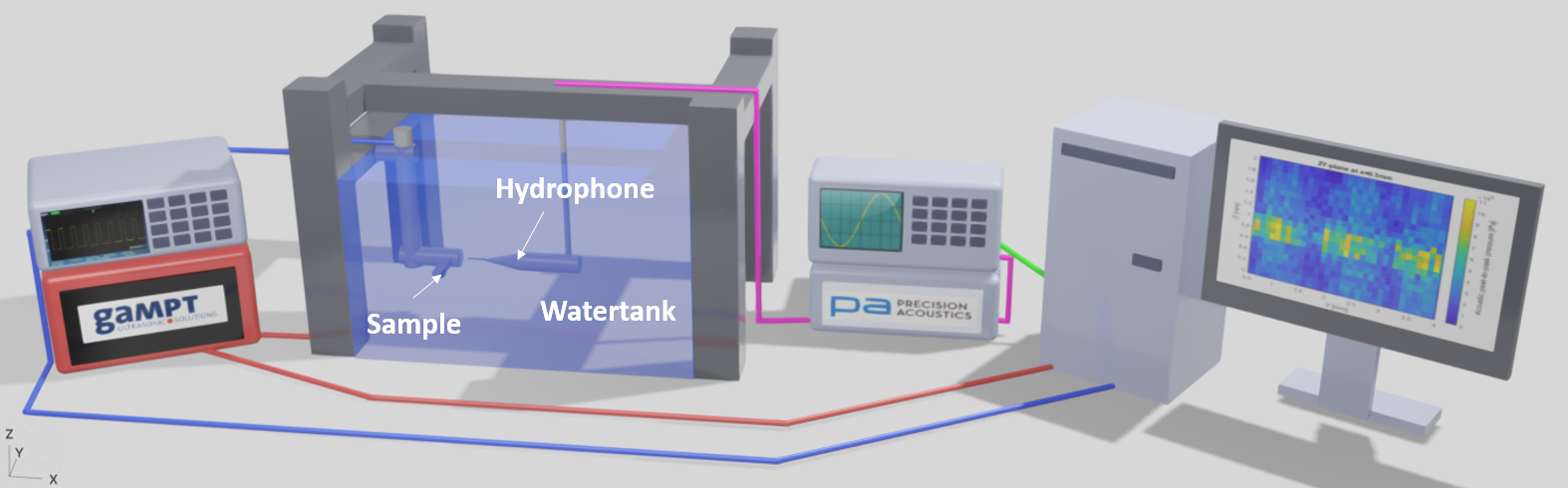}
    \caption{The measurement setup from left to right: the function generator, the 3D-axis motorized stage, the watertank with a hydrophone, the oscilloscope, the hydrophone amplifier, the computer with software.}
    \label{fig:setup}
\end{figure*} 
\subsection{Wafer-level microfabrication}
\label{subsec:fab}
The processing steps for the proposed wafer-level microfabrication process can be found in Fig.~\ref{fig:processing}. A \mbox{300~$\mu m$-thick} double-sided polished 100~mm diameter p-type silicon wafer has been used as a starting material (Fig.~\ref{fig:processing}A). On top of the wafer 1~$\mu m$ of Plasma-Enhanced Chemical Vapor Deposition (PECVD) oxide is deposited at 400$^\circ$C for insulation and as a landing layer for deep reactive ion etching (DRIE) from the backside of the wafer, required later in the process. On top of this layer, a metal interconnect layer of 1~$\mu m$-thick \textit{AlSi} (99\%/1\%) is sputtered at 50$^\circ$C (Fig.~\ref{fig:processing}B). \textit{AlSi} (99\%/1\%) has been used due to its high conductivity, low cost, and availability. This metal layer is patterned using a 2.1~$\mu m$-thick positive photoresist (SPR3012, Shipley) as a soft mask and is etched using HBr/Cl$_2$-based dry etching processes (Fig.~\ref{fig:processing}C). \\
Next, a 4~$\mu m$-thick PECVD $SiO_2$ layer at 400$^\circ$ is deposited at the backside as a hard mask (Fig.~\ref{fig:processing}D). The PECVD oxide layer at the backside is opened using a fully dry etch step (Fig.~\ref{fig:processing}E). For this etch step a 3.1~$\mu m$-thick positive photoresist (SPR3012, Shipley) as soft mask has been used. Afterward, the bulk silicon of the wafer is etched till the $SiO_2$ layer at the top side using DRIE (Fig.~\ref{fig:processing}F). This creates a 1~$\mu m$-thick $SiO_2$ membrane in between rigid, silicon islands. This $SiO_2$ membrane serves as support during the parylene-C coating later on in the process. \\
Next, the wafer is diced in a 2-phase dicing process (Fig.~\ref{fig:processing}G) using the dicer (DAD3221, Disco). In the first phase, the wafer is attached with the top side to an ultraviolet-sensitive dicing foil and the wafer is diced into several larger pieces of around 3~$mm^2$ by 3~$mm^2$. After releasing, each piece is individually diced into separate devices. For this phase, an acetone-sensitive dicing foil is used, since the devices can be self-released from the foil using acetone, thus preserving the thin $SiO_2$ membrane. To maintain the thin $SiO_2$ membrane during dicing, the dicing speed is set to 1~mm/s and a thin silicon edge (10~$\mu m$) is preserved, which does not interfere during the bending process.  \\
Commercial PZT-5H 8~MHz sheets from piezo.com are used for the PZT-based US transducers. For conduction purposes, a 30~$\mu m$-thick anisotropic conductive film (ACF, ARclad 9032-70) is attached to one side of the PZT-5H sheet before dicing. The other side of the PZT has a 0.1~$\mu m$ sputtered nickel layer (Fig.~\ref{fig:design}). The PZT-sheet with ACF is diced into the sizes presented in Table~\ref{tab:design_param}. With the pick-and-place tool (T-300 bonder, Accelonix), the PZTs are placed on the metal contact rings at the silicon substrate (Fig.~\ref{fig:processing}H). Despite the beamsplitter being used for alignment, some spatial variation exists. \\
\indent The top connection between the PZT-based US transducers is made with 50~$\mu m$-thick tungsten wire which is connected using a layer of silver conductive paste (42469, Thermofischer). After curing, a layer of conductive epoxy (EPOTEK-H20E, Epoxy technology) is applied. This gives a mechanically robust and electrically conductive connection. To avoid, mechanical interference of the wires during the curvature of the device, each PZT and contact pad is individually connected with a tungsten wire (Fig.~\ref{fig:processing}I). Afterward, the wires of the three contact pads are bundled, likewise the wires of the three PZT-based US transducers. In this way, two connections, one for the ground and one for the signal, are available during the measurements. After the attachment of the tungsten wire, the device is encapsulated (Fig.~\ref{fig:processing}J) using a 5~$\mu m$-thick parylene-C coating (PDS 2010, Specialty coating systems). Parylene-C is known due to its conformal coating properties and high chemical inertness \cite{Ortigoza2018}. In addition, it is mechanically flexible and therefore used as the flexible interconnect between the silicon islands in the island-bridge design (Fig.~\ref{fig:design}). 
A micrograph of a single device after step H (Fig.~\ref{fig:processing}H) is shown in Fig.~\ref{fig:fig_devices}A. Fig.~\ref{fig:fig_devices}B shows a micrograph of a device with attached wires.
 \begin{figure}[!b]
    \centering
    \includegraphics[width=0.49\textwidth]{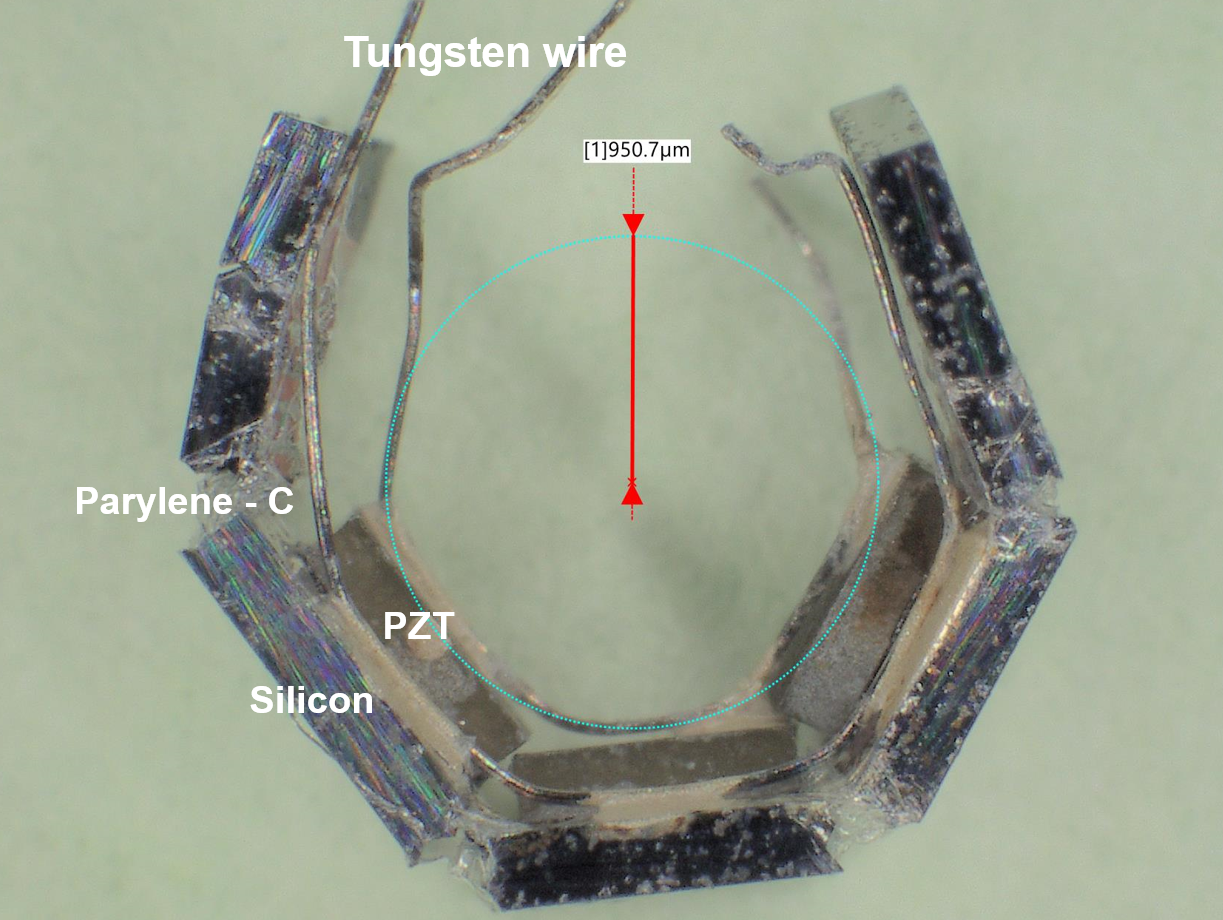}
    \caption{The curvature of the island-bridge structure.}
    \label{fig:curved}
\end{figure} 
\section{Characterization}
\label{sec:char}
\begin{figure*}[t!]
    \centering
    \includegraphics[width=\textwidth]{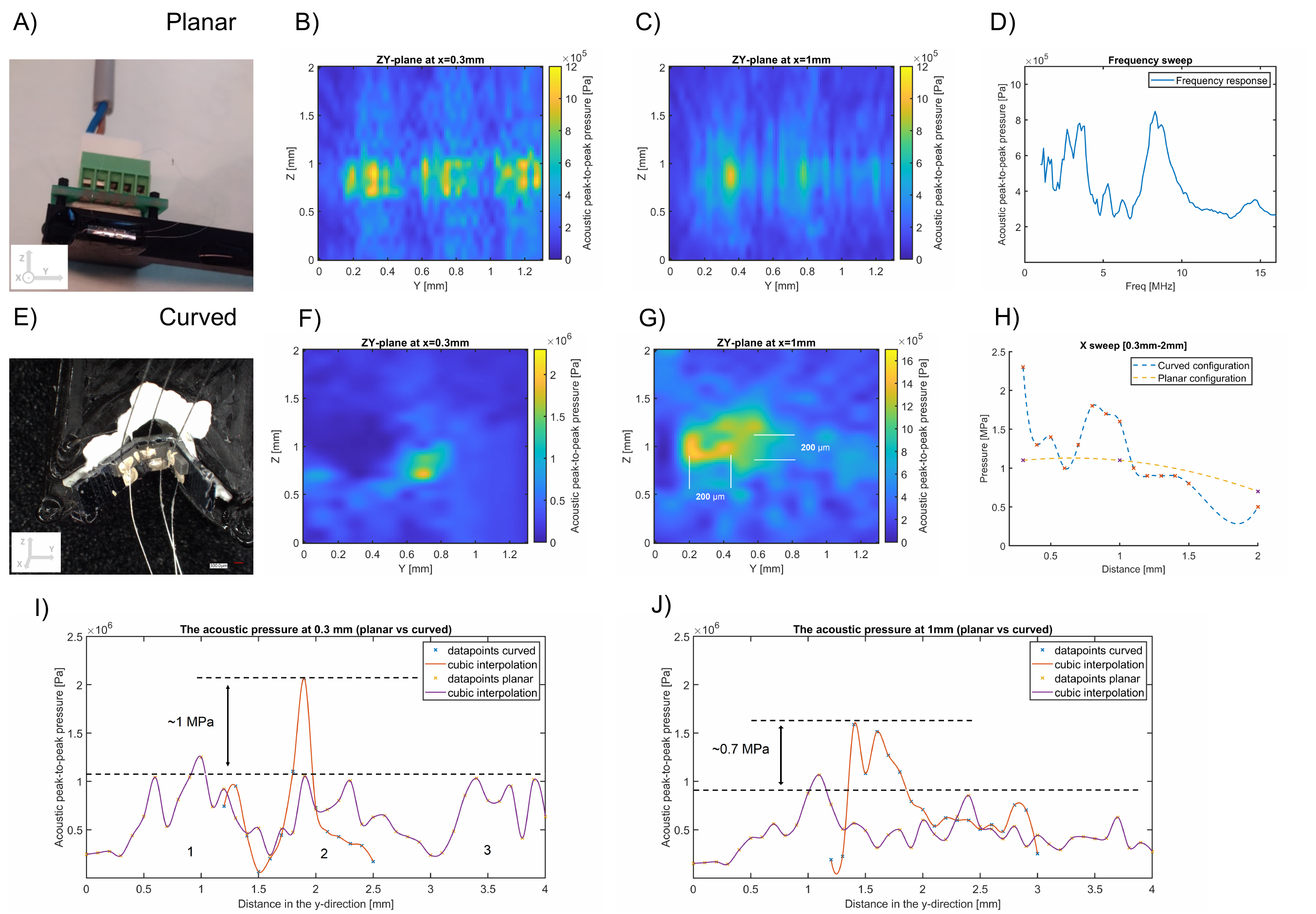}
    \caption{The measurements A) shows a photograph of the 3D-printed holder with the device in a planar configuration B) and C) show the acoustic profile for an x-distance parallel to the surface of the PZT-based US transducers of 0.3 and 1~mm respectively D) shows the frequency response of one of the PZT-based US transducers from the planar configuration E) shows a micrograph of the design being curved in the 3D-printed holder F) and G) show the acoustic profile for an x-distance parallel to the surface of the middle PZT-based US transducer of 0.3 and 1~mm respectively H) shows the pressure profiles of both the planar and curved configuration I) shows the pressure profiles along the white dashed lines in the acoustic profile of the curved and planar configuration at and x-distance of 0.3~mm J) shows the pressure profiles along the white dashed lines in the acoustic profile of the curved and planar configuration at and x-distance of 1~mm.}
    \label{fig:meas}
\end{figure*}

The characterization has been done to show the impact of curvature on the focal spot. 
For the measurements, a device has been measured in a planar and curved configuration. The measurements are done in a water tank in which the device is fixed in a 3D-printed holder. The measurement setup is shown in Fig.~\ref{fig:setup}. A function generator (DG4202, RIGOL) drives the device, generating a 10~$V_{pp}$, 30 pulses, 8.3~MHz, 1~ms period burst. 
The US pressure is measured using a fibre-optic hydrophone (FSV2-5580-10, Precision Acoustics) which is put into position with a \mbox{3D-axis} motorized stage (SFS630, GAMPT soundfield scanning drive). The fibre-optic hydrophone is connected to the \mbox{fibre-optic} hydrophone system (FOHSv2, Precision Acoustics) and the signal is read out with an oscilloscope (DSO-X 3032A, Agilent Technologies). The oscilloscope, function generator and \mbox{3D-axis} motorized stage can be controlled using a \mbox{Matlab-GUI} on a computer. The hydrophone has a sensitivity of 268~mV/MPa at a frequency of 8~MHz. Linear interpolation gives a sensitivity of 281~mV/MPa for 8.4~MHz. The 3D-printed holder for the measurements in the planar configuration can be seen in Fig.~\ref{fig:meas}A. A small custom-made PCB board is attached to connect the device to the oscilloscope connectors. \\
Before the acoustic measurements, a frequency sweep (from 1 to 16~MHz) was applied to the PZT-based US transducers to obtain the resonance frequency of the device (Fig.~\ref{fig:meas}D). For this measurement, a resonance frequency of 8.3~MHz was obtained which is used as the driving frequency for the function generator. The acoustic profiles are measured in a zy-plane parallel to the front of the device for different distances in the x-direction. The data is post-processed using a cubic interpolation method with 100 in between points at both axes. The scans for the 0.3~mm (near-field) and 1~mm (focal spot) distance can be found in Fig.~\ref{fig:meas}B and Fig.~\ref{fig:meas}C respectively. Each PZT-based US transducer has its own acoustic profile and some profile distortion is visible. The acoustic peak pressure varies among the PZT-based US transducers from 1.1~MPa to 700~kPa (Fig.~\ref{fig:meas}C) gaining 900~kPa on average. The focal spot of a single PZT-based US transducer has a size of around 100~$\mu m$ by 200~$\mu m$. The $S_{tx}$ reaches 110~kPa/V. \\
A fully curved device can be observed in Fig.~\ref{fig:curved} with an inner radius of 0.95~mm. A topview of the setup of the curved sample in the water tank is given in Fig.~\ref{fig:meas}E. The 3D-printed holder contains a half-circle which, together with the device, has an inner diameter of 2~mm. The device is pushed inside the half-circle into a thin layer of glue pad that holds the device in a half-curved position. The acoustic profiles are scanned in the same way as for the planar configuration. The scans for 0.3~mm and 1~mm in the x-direction are shown in Fig.~\ref{fig:meas}F and Fig.~\ref{fig:meas}G respectively. The focal spot size is around 200~$\mu m$ by 200~$\mu m$ and is slightly larger than the simulations (Fig.~\ref{fig:meas}G). \\
The focal pressure magnitude in the curved configuration is increased by 0.7~MPa compared to the average focal pressure magnitude of the single PZT-based US transducer in the planar configuration (Fig.~\ref{fig:meas}I and Fig.~\ref{fig:meas}J). Having a peak focal pressure of 1.6~MPa results in a $S_{tx}$ of 160~kPa/V. The pressure profiles of Fig.~\ref{fig:meas}I and Fig.~\ref{fig:meas}J are obtained from the cross sections of Figures~\ref{fig:meas}B and ~\ref{fig:meas}C at Z=0.9~mm, from Figure~\ref{fig:meas}F at Z=0.7~mm, and from Figure~\ref{fig:meas}G at Z=1~mm. For the planar pressure profile line in Figure~\ref{fig:meas}I, the locations of the PZT-based US transducers in the graph are indicated with the numbers 1, 2, and 3. In Figure~\ref{fig:meas}H, the maximum focal pressures for each measurement in the curved and planar configuration are combined in a distance plot. The crosses indicate the maximum values, whereas the dashed line is an interpolation.
\section{Discussion}
\label{sec:disc}
The simulations show that the focal region of the cuff implant design has grating lobes resulting in high-intensity and high-pressure areas around the focal spot (Figure~\ref{fig:sim1}F). This can potentially modulate unwanted regions in the VN. Beam steering could be implemented to reduce these grating lobes and target the nerve more specifically \cite{kawasaki2021}. 
Comparing the results with the simulations, it can be observed that the resonance frequency is well preserved after the fabrication of the device. The simulated resonance frequency is 8.4~MHz whereas the measured resonance frequency is 8.3~MHz. The distortion and harmonics at lower frequencies could be explained by the loading of the PZT-based US transducers due to the attachment of the tungsten wire changing the frequency behavior. Another reason might be the partial detachment of the PZT from the substrate as research shows that that can induce harmonics \cite{Nitesh2014}. The detachment of the PZT from the substrate might be a consequence of poor adhesion to the ACF, placement variations of the PZTs, mechanical vibrations during operation or corrosion of the metal tracks due to water inlet via microcracks in the \mbox{parylene-C} encapsulation. \\
Moreover, the measurement in the curved configuration shows that a local maximum exists in the near-field that has a higher pressure magnitude than the focal spot (Figure~\ref{fig:meas}F). This could potentially result in unwanted VN areas being modulated. These near-field maxima are highly dependent on the medium and thus difficult to model \cite{Gougheri2019}. They could for instance be reduced by implementing beam-steering \cite{Costa2021} or improving the curvature of the device during the measurements. 
To reduce the variations and distortions, some assembly steps could be fine-tuned. The process parameters of the \mbox{pick-and-place} of the PZTs can be refined and it can be automated, as this is a standard packaging step. This will result in more precise PZT placement. Moreover, it could be transformed into a top and backside dicing approach at which the dicing determines the alignment of the PZTs \cite{Costa2021}. The manual attachment of the tungsten wire could be replaced by adding an evaporated or sputtered top metal plane on top of the PZT-based US transducers which will hypothetically reduce the distortion in the US profile. \\
\indent Another reason for the difference between the pressure profile in the simulations and the measured profile is the simplifications and idealities in the simulation model. In the simulation, only the PZT and a perfectly cylindrical parylene-C ring are taken into account. In reality, fabrication non-idealities, island-bridge instead of pure parylene-C, manual variations during PZT placement, and the attachment of the tungsten wire, degrade the performance of the PZT-based US transducer, leading to a different pressure profile. Moreover, the distortion of the focal spot might result from the non-perfect curvature and PZT placement variations. Due to small misalignments and tilting of the sample in the water tank during measurements in the planar configuration, there is a difference between the measured acoustic pressures among the PZT-based US transducers. \\
\indent The island-bridge structure with silicon islands and parylene-C interconnects gives flexibility and the ability of the device to have a curvature of 2~mm in diameter (Fig.~\ref{fig:curved}). However, as parylene-C is naturally brittle, it is still a vulnerability and the device should be handled with care. The robustness could be improved by increasing the layer thickness or creating a multilayer on top which protects the underlying \mbox{parylene-C} layer. However, this might affect the acoustic performance as the attenuation could increase, depending on the layer thickness, material properties, and the driving frequency. A biocompatible, transparent and cuff-implant-suitable alternative is Polydimethylsiloxane (PDMS) \cite{Miranda2021}. Research shows that this material can be used in combination with parylene-C as an encapsulation layer \cite{Babaroud2020, Bakhshaee2021}. \\
\indent For the metal layer, \textit{AlSi} (99\%/1\%) is used. However, during measurements, device failure occurs. This is likely due to the non-optimized adhesion between the parylene-C and the metal tracks and the high water vapor transmission rate (WVTR) of parylene-C \cite{Nanbakhsh2019}. Post-treatment of the metal layer or replacing it by another more inert metal could improve the robustness of the metal layer\cite{Mahmood2022}. Moreover, a multilayer encapsulation might increase the robustness \cite{Pak2022} as well. Another advantage is that a multilayer encapsulation could be used for acoustic matching. A matching layer increases the acoustic power transfer between two non-matching media (PZT and water) and will increase the acoustic output pressure. Besides a backing layer, a matching layer could be included in the wafer-level microfabrication process as well \cite{Rathod2020}. The multi-layer polymer-metal structure for acoustic matching might be promising as parylene-C can still be used as an encapsulation layer \cite{Fei2015, Toda2012, Yang2019}.  \\
\indent To increase the acoustic output pressure even more, \mbox{PZT-5H} piezoelectric material could be replaced by the \mbox{PMN-PT} piezoelectric material as it has better electromechanical properties \cite{Kim2010}. Moreover, beam steering could be implemented by dicing each individual PZT-based US transducer into a 2D-phased array \cite{Costa2021}. This opens the potential to target the VN at various locations within the radius of the cuff implant. Another benefit of beam steering is the ability to compensate for mechanical deformation. \\
\indent The measurements show that the fabricated device in curved configuration has a high $S_{tx}$, indicating that per applied driving voltage to the device the output pressure increases with that amount \eqref{eq:sensitivity}.
\begin{equation}
    P_{out} = V_{driving}*S_{tx}
\label{eq:sensitivity}
\end{equation}
For a driving voltage of 10~V used in this study, the output pressure is $P_{out} = 160 * 10 = 1.6~MPa$ using the $S_{tx}$ of 160 kPa/V. For comparison, CMUTs designed for VN neuromodulation have a $S_{tx}$ of 68 kPa/V \cite{kawasaki2021}. In addition, a body-conformal active ultrasound patch presented by Pashaei et al shows a $S_{tx}$ of 80 kPa/V \cite{Pashaei2020}. This means that for similar driving voltages, significantly higher output pressures are obtained with this proposed design.

\section{Conclusion}
\label{sec:concl}
This paper proposes a 1~mm by 7~mm wafer-level microfabricated, island-bridge cuff implant with an inner diameter of 2~mm. COMSOL Multiphysics simulations have been performed to investigate the effect of the number of PZT-based US transducers and to verify the design. The wafer-level microfabrication and assembly consist of standardized and scalable process steps. \\
 The device is driven at 8.3~MHz and has a focal length of 1~mm. Three commercial PZT-5H US transducers are integrated generating 0.9~MPa on average at the focal spot for each individual PZT-based US transducer in a planar configuration ($S_{tx}$ = 110~kPa/V) whereas, 1.6~MPa is generated at the focal spot in curved configuration ($S_{tx}$ = 160~kPa/V). The focal spot of the curved cuff implant is around 200~$\mu m$ by 200~$\mu m$. \\
The measurements show the potential of a cuff-shape design with a PZT-based US transducer array as the output focal pressure is increased by at least 45\% (taking the peak pressures at the focal spot for both the planar (1.1~MPa) and curved (1.6~MPa) configuration) compared to the measured focal pressures of the single PZT-based US transducers in the planar configuration.  
In conclusion, the integration of PZT-based US transducers in a \mbox{cuff-shaped} design opens a new path towards a technique for \mbox{high-precision} VNS. 
\section*{Acknowledgment}
 The authors highly appreciated the support of the staff of the Else Kooi Lab at Delft University of Technology. 


\end{document}